\RequirePackage{fix-cm}

\documentclass[numbook]{svjour3} 
\journalname{Empirical Software Engineering}
\counterwithin*{section}{part}
\usepackage[a4paper,bindingoffset=0mm,left=30mm,right=30mm,top=20mm,bottom=20mm]{geometry}

\usepackage{array}
\newcolumntype{L}{>{\arraybackslash}m{12cm}}

\newcolumntype{P}{>{\arraybackslash}m{6cm}}
\newcolumntype{C}{>{\centering\arraybackslash}m{12cm}}

\def\BibTeX{{\rm B\kern-.05em{\sc i\kern-.025em b}\kern-.08em
    T\kern-.1667em\lower.7ex\hbox{E}\kern-.125emX}}

\makeatletter
\renewcommand\paragraph{\@startsection{paragraph}{4}{\z@}%
  {0.5ex \@plus 1ex \@minus .2ex}%
  {-0.5em}%
  {\normalfont\normalsize\itshape}}
\makeatother

\usepackage{cite}

\usepackage{amsmath,amssymb,amsfonts}
\usepackage{algorithmic}
\usepackage{graphicx}
\usepackage{textcomp}
\usepackage{xcolor}
\def\BibTeX{{\rm B\kern-.05em{\sc i\kern-.025em b}\kern-.08em
    T\kern-.1667em\lower.7ex\hbox{E}\kern-.125emX}}

\usepackage{cleveref}

\usepackage[framemethod=TikZ]{mdframed} % Ensure you import this in your preamble
\usepackage[comma,authoryear]{natbib}
\usepackage{booktabs}
\usepackage[most]{tcolorbox}
\usepackage{xspace}
\usepackage{caption}
\usepackage{stfloats}
\usepackage{balance}
\usepackage{tabularx} % Add to preamble

\usepackage{url}
\newcommand{\takeaway}[2]{%
    \vspace{8pt}%  % Adjust this value as needed
    \begin{tcolorbox}[
        enhanced,
        colback=gray!10,
        colframe=black,
        boxrule=0.75pt,
        arc=3pt,
        top=6pt,
        bottom=6pt,
        left=6pt,
        right=6pt,
        boxsep=0pt,
        before={\noindent},
        after={\par\vspace{4pt}},  % Adjust this too
        grow to left by=0pt,
        grow to right by=0pt
    ]
        \textbf{#1:} #2
    \end{tcolorbox}
}

\def\Producers{Producers\xspace}

\def\consumers{consumers\xspace}
\def\producers{producers\xspace}

\begin{document}

\title{When Model Release Meets Model Reuse: Producer-Consumer Misalignment in Hugging Face\\

\author{Adekunle Ajibode  \and
        Oussama Ben Sghaier 
        \and Bram Adams
        \and Ahmed E. Hassan}
}

\institute{Adekunle Ajibode \at
              School of Computing, Queen’s University, Kingston, ON, Canada\\
              \email{ajibode.a@queensu.ca}%  \\
           \and
           Oussama Ben Sghaier \at
              School of Computing, Queen’s University, Kingston, ON, Canada\\
              \email{oussama.sghaier@queensu.ca}
              \and
           Bram Adams \at
               School of Computing, Queen’s University, Kingston, ON, Canada\\
              \email{bram.adams@queensu.ca}
              \and
           Ahmed E. Hassan \at
               School of Computing, Queen’s University, Kingston, ON, Canada\\
              \email{hassan@queensu.ca}}

\date{Received: date / Accepted: date}

\maketitle

\begin{abstract}
Pre-trained Language Models (PTLMs) are increasingly reused as dependencies in modern software systems, even though prior work has documented persistent structural problems in PTLM supply chains, such as inconsistent release practices, incomplete metadata, and divergence between Hugging Face and GitHub repositories. One previously unexplored angle on these problems is that misalignment between PTLM producers' release practices and consumers' model reuse needs may explain several of these challenges, yet the human expectations, processes and interactions underneath this misalignment have never been studied. As such, we conducted the first dual-perspective survey of 50 Hugging Face producers and 95 GitHub consumers to examine producer-consumer misalignments across four AI supply chain dimensions: model discovery, documentation practices, lineage tracing, and model governance adoption. We find that,  although producers and consumers rely on the same documentation artifacts, they disagree on where critical metadata should be recorded. Even though 27.9\% of producers and 31.4\% of consumers trace lineage beyond the immediate parent model, producers primarily trace model lineage for provenance and reproducibility, whereas consumers are mainly driven by quality and reliability concerns. Furthermore, producers believe that governance mechanisms that streamline model release would have the greatest positive impact, whereas consumers prioritize documentation and dependency transparency. These findings highlight opportunities to improve model documentation conventions, lineage visibility, and governance support in PTLM supply chains.
\end{abstract}


\section{Introduction}\label{introduction}
Pre-trained Language Models (PTLMs) have driven a paradigm shift in software development, demonstrating increasingly powerful capabilities from code generation to autonomous reasoning~\citep{zhao2023survey, min2023recent}. PTLMs are increasingly treated as reusable, adaptable components across downstream tasks rather than standalone computational artifacts~\citep{bommasani2021opportunities}. In contrast to traditional code dependencies~\citep{decan2019empirical, yasmin2025software}, PTLMs encode learned behaviors shaped by data and optimization rather than explicit human logic~\citep{bommasani2021opportunities, yasmin2025software}.

PTLM reuse has given rise to a complex AI supply chain, which, in this paper, we define as the end-to-end ecosystem of upstream and downstream actors involved in the publication, discovery, reuse, and governance of pre-trained models for software engineering applications~\citep{jiang2022empirical, jiang2024peatmoss}. Prior work has shown that this supply chain covers a highly diverse socio-technical ecosystem of actors, artifacts, and processes, including model providers, model registries, model inference engines, and third-party libraries~\citep{hopkins2025ai}.   

The AI supply chain introduces distinct engineering challenges in model artifact publication, configuration, and downstream model reuse (e.g., fine-tuning)~\citep{banyongrakkul2025release}. Recent empirical studies reveal persistent structural problems: poor model naming and versioning practices~\citep{ajibode2025towards,jiang2025see}, poor coordination between AI and software supply chains~\citep{adekunle2025synchronization,jannelli2026agentic}, and incomplete metadata that complicates model lineage verification~\citep{stalnaker2025empirical,yang2024navigating,garcia2026ethical}. Unlike traditional software engineering where provenance tracking and compliance are automated via Software Bills of Materials (SBOMs), emerging governance mechanisms like AI Bills of Materials (AIBOMs) have not yet gathered large-scale adoption and tool support.

Resolving these model reuse challenges requires a clear understanding of the human ecosystem driving this supply chain, which is anchored by two primary actors, ``producers", who create, train, or publish models on platforms like Hugging Face, and ``consumers", who reuse these models in downstream applications. However, while existing studies have started to document the observable symptoms of AI supply chain friction, to our knowledge, no empirical study has directly surveyed and contrasted the experiences of both model producers and consumers to understand the human decisions and workflows underneath the above symptoms.

Therefore, this paper aims to understand how producers and consumers interact across the AI supply chain and where potential misalignments arise. We surveyed 50 PTLM producers on Hugging Face and 95 PTLM consumers on GitHub across four AI supply chain dimensions: PTLM discovery and pre-adoption trust, PTLM documentation practices, supply chain lineage visibility, and supply chain governance trade-offs. These surveys address these four research questions:

\noindent\textbf{RQ1: How do producers and consumers discover PTLMs and what are the risks of model reuse?}

\noindent\textbf{Motivation:} Prior research has studied PTLM supply chains using artifact-based signals such as download counts, repository metadata, and dependency relationships~\citep{jiang2022empirical, taraghi2024deep}, but provides limited insight into how practitioners discover models, assess trustworthiness, and mitigate adoption risks. This RQ examines these practices from both \producers and \consumers perspectives.

\noindent\textbf{Result:} While both producers and consumers prioritize model size during discovery, the former also focus on knowledge of the (base) model a model is derived from, while the latter focus on model documentation quality. However, producers have to deal with hidden biases in model outputs and data contamination issues, while consumers face incomplete or unclear documentation and reproducibility risks.

\noindent\textbf{RQ2: How do producers document PTLMs, and how do these practices align with consumer information needs?}

\noindent\textbf{Motivation:} Documentation artifacts such as ``Model Cards'' aim to improve transparency and accountability~\citep{mitchell2019model}, yet prior work has reported inconsistencies and omissions~\citep{yang2024navigating}, and proposed improvements~\citep{bhat2023aspirations}. This RQ examines practitioner documentation practices, barriers, and alignment with consumer information needs.

\noindent\textbf{Result:} Time pressure and technical complexity contribute to incomplete documentation, creating systematic metadata-location misalignments with consumer expectations. Consumers compensate through cross-platform triangulation, empirical testing, and community signals.

\noindent\textbf{RQ3: What practical strategies do producers and consumers use to understand and trace model lineage across the AI supply chain, and what limitations or challenges do these strategies exhibit?}

\noindent\textbf{Motivation:} Prior work has used repository mining to reconstruct model lineage, that is, the chain of parent-child dependencies that connects a model to earlier models it was derived from using techniques such as fine-tuning, quantization, or merging~\citep{jiang2024peatmoss}. Although lineage is one aspect of model provenance (i.e., a model's entire origin and history), little is known about how practitioners trace lineage in practice or respond to incomplete dependency information.

\noindent\textbf{Result:} Producers and consumers use documentation inspection, platform tools, version tracking, and multi-source verification to trace lineage, but face fragmented model registries and undocumented base models (i.e., root models).

\noindent\textbf{RQ4: What governance mechanisms do \producers and \consumers prioritize, and what barriers block them?}

\noindent\textbf{Motivation:} Building on the definition of AI governance as ``a system of rules, practices, processes, and technological tools that ensure AI aligns with organizational objectives, legal requirements, and ethical principles~\citep{samuli2022ai}", we use the term \emph{governance mechanisms} to refer to the individual processes, policies, and technical tools that improve transparency, traceability, and accountability in AI supply chains. Emerging proposals such as SPDX 3.0's AI bills of material (AIBOM) aim to improve governance by documenting model provenance, training data, and dependencies~\citep{kale2023provenance, duan2024modelgo}, yet evidence on practitioner perceptions remains limited.

\noindent\textbf{Result:} Producers prioritize mechanisms that streamline model release and management, whereas consumers prioritize transparency and model selection. Both are hindered by model reuse and workflow compatibility challenges.

This paper makes the following contributions: 
\begin{itemize} 
    \item The first dual-perspective empirical analysis mapping the practices, constraints, and risk perceptions of model \producers against the reuse requirements of model \consumers. 
    \item Identification of producer-consumer misalignments in model documentation and metadata, insights into lineage tracing practices and visibility barriers, and quantification of governance adoption challenges due to tooling deficiencies and workflow overhead. 
    \item A replication package including quantitative analysis scripts, survey instruments, and qualitative codebook~\citep{anonymous2026replication}. 
\end{itemize}

\vspace{-10pt}
\section{Related Work}\label{related_work}

\subsection{Reuse Challenges for Machine Learning Components}
Prior empirical studies on challenges for machine learning components show that machine learning (ML) model reuse is common but introduces maintenance risks similar to code cloning~\citep{sens2025large}, poor modularity and limited testing automation~\citep{nahar2025product}, cascading update defects~\citep{dilhara2021understanding}, and scalability challenges~\citep{shivashankar2025maintainability}. In the context of PTLMs, Taraghi~et~al.~\citep{taraghi2024deep} analyzed model reuse trends on Hugging Face, showing that model architectural changes introduce reuse barriers. Casta{\~n}o~et~al.~\citep{castano2024machine} characterized how machine learning models evolve over time and affect downstream stability. However, these studies focus on technical aspects and decisions rather than human processes underlying model discovery and reuse.

\subsection{AI Supply Chain Transparency and Practitioner Perspectives}

Prior research has examined AI supply chain transparency, compliance, and risk using repository mining of dependency graphs~\citep{jiang2024peatmoss}, licensing frameworks~\citep{duan2024modelgo}, issue resolution times~\citep{banyongrakkul2025release}, and security threats~\citep{wang2025large, hu2025large}. These studies have reported concrete anomalies, including vulnerability propagation from upstream to downstream consumers~\citep{jiang2022empirical}, legal verification gaps due to missing dependency declarations~\citep{stalnaker2025empirical}, synchronization inconsistencies between Hugging Face and GitHub repositories~\citep{adekunle2025synchronization}, and metadata inconsistencies in naming, versioning, and licensing~\citep{jiang2023naming, taraghi2024deep, castano2024machine, ajibode2025towards}. However, because these approaches rely on external model artifact properties, they provide limited insight into the motivations, constraints, and decision-making processes underlying these observed issues. 

To complement this perspective, our study adopts a dual-perspective survey design, following prior software engineering work on supply chain roles and workflows~\citep{vassallo2019automated, wan2019does}. Vassallo et al.~\citep{vassallo2019automated} surveyed CI practitioners to uncover misalignments between developer practices and infrastructure capabilities, while Wan et al.~\citep{wan2019does} surveyed ML practitioners to understand how ML changes software development practices and reveals gaps between expectations and reality. We extend this approach to the AI supply chain, focusing on model-centric dependencies where learned behaviors replace explicit logic. By surveying producers and consumers directly rather than relying solely on repository traces, we expose misalignments across model discovery, documentation, lineage tracing, and governance mechanisms that mining studies cannot reveal.

\subsection{Model Documentation and Transparency Artifacts}
To improve machine learning model transparency, researchers proposed frameworks like Model Cards~\citep{mitchell2019model}, Datasheets~\citep{gebru2021datasheets}, and AIBOMs~\citep{kale2023provenance,d2025aloha}. However, empirical audits and practitioner studies show that model documentation remains incomplete, unstandardized, and difficult to maintain in practice~\citep{yang2024navigating,mamirov2025ai,garcia2026ethical,bhat2023aspirations}. Prior work highlights poor version control~\citep{toma2024exploratory, ajibode2025towards}, license drift across hubs~\citep{jewitt2025hugging}, and sparse ethics documentation on Hugging Face~\citep{oreamuno2024state}. This metadata sparsity creates a transparency debt, e.g., AIBOM standards exist in theory, but in practice they are hard to generate or audit because the required data is missing~\citep{rajbahadur2025building}. While frameworks exist to track model lineage quality~\citep{gong2023survey} or provide supply chain datasets like PeaTMOSS~\citep{jiang2024peatmoss}, there remains a scarcity of research investigating producer-consumer model alignment across four AI supply chain dimensions: model discovery, documentation practices, lineage tracing, and governance adoption.

\section{Methodology} \label{sec:methodology}
This section outlines our empirical methodology to address the research questions as summarized in \Cref{introduction}.

\begin{figure*}[htbp]
\centering
\includegraphics[width=\textwidth]{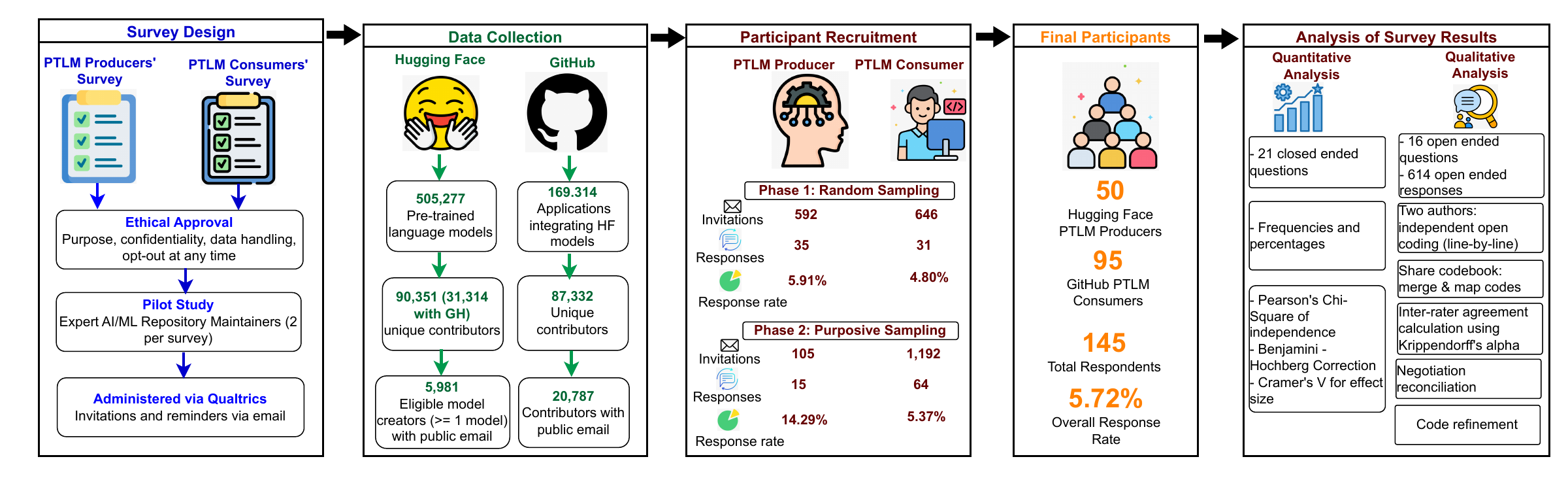}
\captionsetup{skip=-9pt}
%\captionsetup{skip=0pt}
\caption{Overview of the study methodology.}
\label{methodology}
\end{figure*}

\vspace{-4pt}

\subsection{Survey Design and Pilot}

The study received ethics approval from our university prior to data collection. To capture dual perspectives across the AI supply chain, we developed two survey instruments: one for Hugging Face producers with at least one published model~\citep{laufer2025anatomy}, and one for GitHub consumers who had reused one or more Hugging Face models in their applications.

Survey questions were derived from our research questions and prior literature. The first author reviewed empirical studies on AI supply chains and model reuse (2019--2025) from ICSE, ESEC/FSE, TOSEM, EMSE, FAccT, CHI, and NeurIPS, covering four dimensions aligned with our RQs: model discovery and trust, documentation practices, lineage and provenance tracing, and governance mechanisms. Discovery questions drew on artifact-based signals, naming and versioning, and model popularity \citep{taraghi2024deep, ajibode2025towards, kadasi2025model}; documentation questions on Model Card incompleteness, README omissions, and sparse metadata \citep{yang2024navigating, bhat2023aspirations, garcia2026ethical, oreamuno2024state, stalnaker2025empirical, mamirov2025ai}; lineage questions on dependency reconstruction, provenance gaps, and derivative relationships such as model fine-tuning, quantization, and merging \citep{jiang2024peatmoss, stalnaker2025empirical, banyongrakkul2025release}; and governance questions on AIBOM, SPDX, and licensing proposals adapted into questions on adoption barriers and perceived usefulness \citep{kale2023provenance, bennet2025implementing, duan2024modelgo}. We also drew on dependency management, SBOM adoption, and supply chain security literature to adapt software supply chain concepts to AI systems \citep{decan2019empirical, srinivasan2025bomfather, seshadri2024omnibor, yasmin2025software}.

Two authors mapped each literature insight to candidate survey questions, refined through four rounds of internal review with all co-authors, while a third author independently audited the mapping for coverage, identifying gaps through additional targeted searches. All authors then reviewed the resulting survey.

The two surveys share 16 core questions on model discovery, trust assessment, documentation use, lineage tracing, and governance, with identical wording across cohorts for direct comparison. The producer survey adds 3 questions on documentation completion rates and release barriers, while the consumer survey adds 2 on coping with missing documentation and selection confidence. As a result, the producer survey contains four demographic and nineteen core questions (eight open-ended, eleven closed-ended), and the consumer survey contains three demographic and eighteen core questions (eight open-ended, ten closed-ended). Open-ended questions allowed participants to elaborate on challenges, decision-making processes, and governance needs.

We piloted each survey with two active industry AI/ML repository maintainers on Hugging Face, refining wording, structure, and length to improve clarity and reduce completion time. Each survey required approximately 15--20 minutes. Full instruments are in our replication package \citep{anonymous2026replication}.

\subsection{Participants Identification}\label{participant_identified}

To identify potential producer participants, we scraped metadata from all 505,277 active language models on Hugging Face as of December 29, 2025, identifying 90,351 unique contributors. From these contributors, we extracted public email addresses from linked GitHub profiles, yielding a sampling frame of 5,981 eligible producers.

To identify potential consumer participants, we mined repository source code on GitHub on February 23, 2026, identifying 169,314 applications containing explicit Hugging Face model imports. This corresponded to 87,332 unique contributors and 20,787 public email profiles.

For both groups, we deduplicated identities by taking the union of all usernames and removing exact string matches on the username field. We did not apply fuzzy matching or other similarity techniques, instead we ensure that the contacted participants cover a diverse set of models or applications, respectively.

\subsection{Recruitment and Demographics}
Email recruitment ran between April 15 and May 15, 2026, utilizing a two-phase strategy. Phase 1 applied probability-based random sampling across all historical contributors identified in \Cref{participant_identified}, targeting a $99\%$ confidence level with a $5\%$ margin of error \citep{foalem2026empirical, jiang2025see}, distributing invitations to $592$ producers ($35$ responses) and $646$ consumers ($31$ responses). Phase 2 utilized purposive sampling to target active producers and consumers with repository contributions within the prior six months, issuing $105$ invitations to producers ($15$ responses) and $1,192$ to consumers ($64$ responses).

Participants completed an informed consent form describing the study purpose, voluntary participation, confidentiality provisions, and data handling procedures before accessing the survey. Non-consenting respondents were exited immediately, and no responses were collected. As the eligibility criteria and target populations remained consistent across both recruitment phases, we combined the responses from Phase~1 and Phase~2, yielding a final cohort of $50$ producers and $95$ consumers (overall response metrics detailed in \Cref{methodology}).

Among producers, the primary roles were Model Integrators ($60.0\%$) and Researchers ($54.0\%$), predominantly split between $1$--$3$ and $4$--$7$ years of experience, with $64.0\%$ publishing $1$--$5$ models annually. A smaller proportion of producers ($6.0\%$) reported $8+$ years of experience. Among consumers, Application Developers ($42.1\%$) and Model Integrators ($37.9\%$) predominated, with $41.1\%$ reporting $1$--$3$ years of domain experience, while $7.4\%$ reported $8+$ years of experience. Complete demographic distributions are provided in \citep{anonymous2026replication}, and limitations regarding AI supply chain reachability are addressed in \Cref{threats}.

\vspace{-7pt}
\subsection{Data Analysis}
\noindent\textbf{Quantitative Analysis:} Closed-ended responses were analyzed via descriptive statistics. To assess statistical significance between the producer and consumer cohorts, we applied Pearson's Chi-squared tests of independence \citep{mchugh2013chi}. For single-choice questions, where respondents selected exactly one option from a set of mutually exclusive choices, we interpreted the chi-squared results using raw $p$-values. For multiple-response questions (where producers/consumers could select multiple options that applied), we disaggregated each option into a separate binary outcome with a $2 \times 2$ contingency table (e.g., selected vs. not selected, by producer vs. consumer). For the latter questions, we corrected for multiple comparisons using Benjamini--Hochberg correction \citep{benjamini1995controlling} to adjust $p$-values for a given question across all Chi-squared tests conducted between the producer and consumer cohorts. We utilized Cramér's $V$ \citep{akoglu2018user} to quantify effect sizes.

\noindent\textbf{Qualitative Analysis:} Open-ended survey questions were analyzed using inductive thematic analysis \citep{braun2006using}. Across both surveys, participants were presented with 16 optional open-ended questions. Since these were not mandatory, responses per question ranged from 5 to 125, with respondents answering an average of 9 out of 16. In total, we manually analyzed 614 open-ended responses.

For each question, two authors independently familiarized themselves with all responses and conducted open coding without communicating with each other, recording codes that concisely capture each response's meaning in separate spreadsheets, yielding two codebooks (a structured list of codes, definitions, and example quotes) per question. The authors then consolidated their codebooks one question at a time, defining and grouping semantically equivalent codes into canonical concepts, mapping each author's original codes to these canonical codes, and meeting in person to negotiate differences in codes and definitions.

After adapting their coding results to the consolidated codebooks, we then measured for each question inter-rater percentage agreement (96.7\%--100\%) and Krippendorff's alpha (0.678--1.000) between both authors, accommodating for varying numbers of codes per response: all alpha values exceeded the 0.667 threshold for tentative conclusions \citep{marzi2024k}. Remaining discrepancies were resolved through negotiated agreement sessions in which both authors revisited the raw survey responses to refine interpretations, with an Excel spreadsheet documenting all changes as an audit trail. Afterwards, to verify code consistency, we randomly selected 10\% of responses, which both authors cross-checked in person against the finalized codebook definitions for each question. Verified codes are summarized in \Cref{all_open_text}. The results of closed questions are not included in that figure and are instead reported within the text. 

\vspace{-5pt}
\section{Results}
\subsection{$RQ_1$: How do producers and consumers discover PTLMs and what are the risks of model reuse?}

\noindent\textbf{While both producers and consumers predominantly start model discovery on Hugging Face, and consider model size their top selection criterion, producers consider base model disclosure the \#2 criterion compared to model documentation quality for consumers.} \Cref{information_importance} shows that producers rated model size (51.2\%) and base model disclosure (48.8\%) as critical, and consumers rated model size (40.3\%) and documentation quality (37.5\%) as critical when selecting models to reuse. In contrast, neither group considered popularity signals or the reputation of the model's organization as important when selecting models, as these options ranked last in their responses, ranging between 8\% and 12\%. When applying a weighted scoring approach that accounts for full importance rankings, documentation and base model disclosure are tied for producers, both at 12.1\%. For consumers, documentation (12.6\%) and model size (12.3\%) emerge as the top factors. Popularity signals again rank lowest for both groups, at 5.3\% for producers and 7.8\% for consumers. Note that consumers relied less on Hugging Face for model discovery (42.1\% vs. 51.2\% for producers), as they complement this platform with external channels such as search engines (8.4\%) and AI chatbots (7.4\%).

\begin{table}[t]
\centering
\caption{Information Importance for Model Selection: Producers vs. Consumers}
\label{information_importance}
\resizebox{\linewidth}{!}{%
\begin{tabular}{lrrrr}
\hline
\textbf{Source} & \textbf{Prod. Crit. (\%)} & \textbf{Cons. Crit. (\%)} & \textbf{Prod. Wtd. (\%)} & \textbf{Cons. Wtd. (\%)} \\ \hline
Documentation & 47.5 & 37.5 & 12.1 & 12.6 \\
Base model disclosure & 48.8 & 25.0 & 12.1 & 9.8 \\
Model size & 51.2 & 40.3 & 12.0 & 12.3 \\
License & 48.8 & 34.7 & 11.7 & 10.5 \\
Pipeline tag or task label & 31.7 & 19.4 & 10.2 & 9.6 \\
Benchmark / leaderboard results & 22.0 & 26.4 & 10.1 & 10.5 \\
Download/usage statistics & 22.0 & 13.9 & 9.3 & 8.8 \\
Presence of example notebooks/code & 9.8 & 22.2 & 8.9 & 9.7 \\
Known organization & 9.8 & 12.5 & 8.2 & 8.5 \\
Number of "likes" or popularity signals & 2.4 & 8.3 & 5.3 & 7.8 \\ \hline
\end{tabular}%
}
\smallskip
\small\\ Note: \textit{Critical Percentage (Crit. \%)} represents the proportion of respondents who explicitly rated a source as ``Critical.'' \textit{Weighted Percentage (Wtd. \%)} reflects a source's relative share of the total points across all options, calculated using scored matrix responses (Critical = 4, Very important = 3, Moderately important = 2, Slightly important = 1, Not important = 0).
\end{table}

\noindent\textbf{PTLM producers and consumers report a diversity of risks when reusing models.} \Cref{reuse_risks} shows that Producers reported hidden biases in model output (32.0\%), model performance not matching claims (32.0\%), and data contamination issues (32.0\%) as their primary concerns. Consumers, in contrast, reported model performance not matching claims (42.1\%), incomplete or unclear documentation (31.6\%), and lack of reproducibility (25.3\%) as their foremost concerns. While license compliance violations were a concern for only 14.0\% of producers, they did impact 23.2\% of consumers. These results show that in practice, participants prioritize different concerns: consumers are more concerned with documentation and reproducibility, while producers fear data contamination alongside biases. The complete breakdown of these priority results is available in our replication package \citep{anonymous2026replication}.

\begin{table}[t]
\centering
\caption{Model Reuse Risks Concerns: Producers vs. Consumers}
\label{reuse_risks}
\begin{tabular}{lrr}
\hline
\textbf{Risk} & \textbf{Producer (\%)} & \textbf{Consumer (\%)} \\ \hline
Hidden biases in model outputs & 32.0 & 20.0 \\
Model performance not matching claims & 32.0 & 42.1 \\
Data contamination / training data issues & 32.0 & 20.0 \\
Lack of reproducibility & 28.0 & 25.3 \\
Security vulnerabilities in model weights or code & 22.0 & 18.9 \\
Dependency on unmaintained upstream models & 22.0 & 9.5 \\
Incomplete or unclear documentation & 18.0 & 31.6 \\
Hidden or undocumented upstream model dependencies & 16.0 & 6.3 \\
License compliance violations & 14.0 & 23.2 \\
Dataset provenance & 12.0 & 9.5 \\
Compute cost unexpectedly high & 10.0 & 15.8 \\ \hline
\end{tabular}
\smallskip
\small\\ Note: Respondents were asked to select their top 3 primary risk concerns when reusing pre-trained models. Percentages represent the proportion of total respondents within each cohort who selected the respective risk factor.
\end{table}

\begin{figure*}[t!]
\centering
\includegraphics[width=\linewidth]{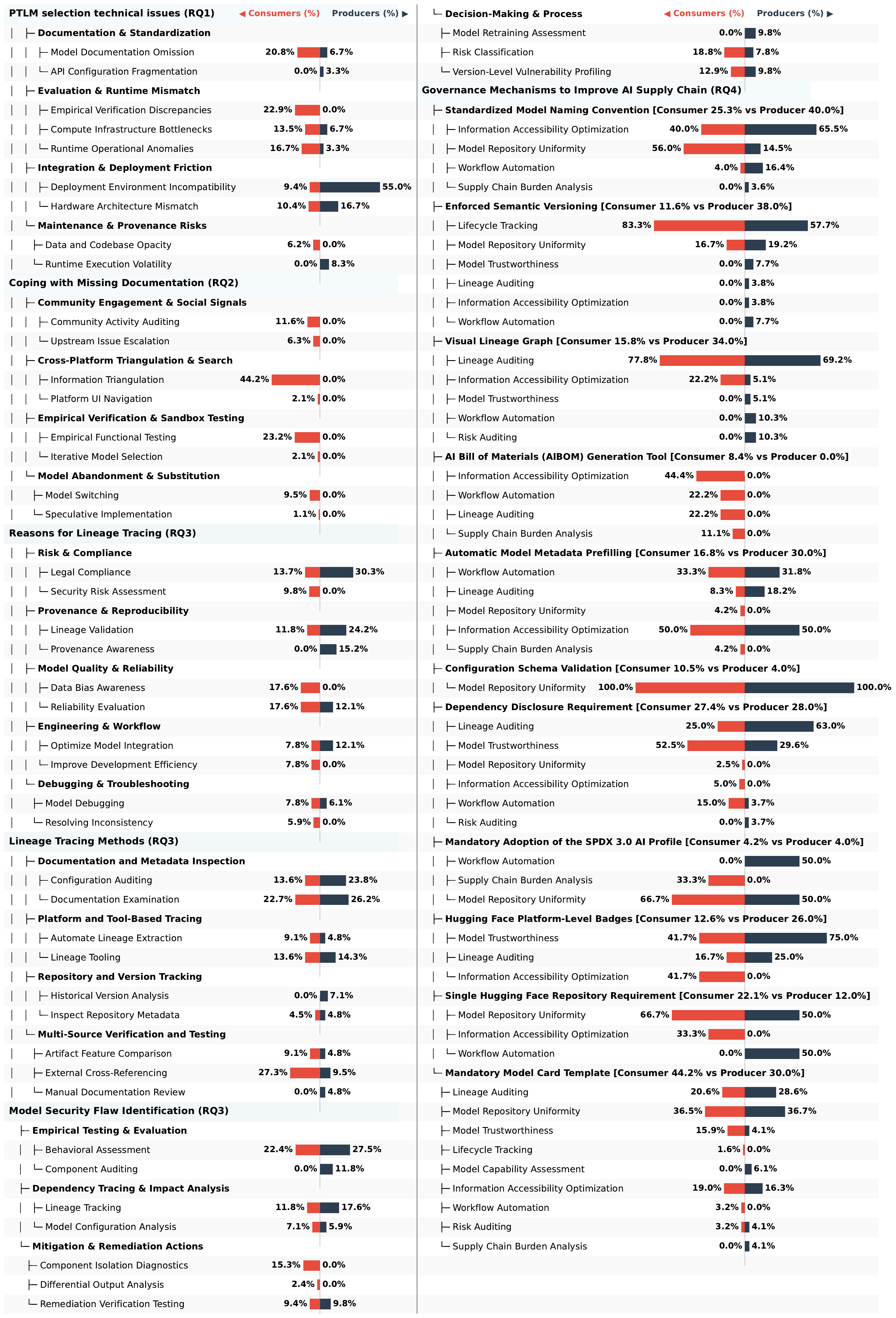}
\caption{\textbf{Codebook distribution of technical issues, reasons for lineage tracing, lineage tracing methodology, flaw identification, mechanisms that can have positive impact in the AI supply chain, and resistance to governance mechanisms (the last open question in RQ4 is in the replication package).}}
\label{all_open_text}
\end{figure*}

Beyond the risks discussed above, open-ended responses revealed additional technical issues that both producers and consumers face when reusing models. As shown in \Cref{all_open_text}, producers predominantly encounter deployment environment incompatibility errors and hardware architecture mismatches, whereas consumers face documentation omissions and performance discrepancies between published claims and local results. For example, one producer explained that \textit{``research labs have the tendency to be too tied to NVIDIA compute, while most models can be trained also on AMD ROCm,"} while a consumer noted that \textit{``I found that many models did not match the performance that the official docs claimed."} While prior work has documented the challenges practitioners face when reusing PTLMs \citep{jiang2023empirical, banyongrakkul2025release}, our dual-perspective survey highlights that producers struggle to make models interact with existing environments (``does it run''), whereas consumers struggle with empirical behavior and accuracy expectations post-reuse (``does it work''). 

\takeaway{Summary of Findings}{\Producers and \consumers share concerns about false performance claims but diverge on other priorities: \producers focus on compatibility, bias, and data quality, whereas \consumers focus on documentation quality, reproducibility, and post-reuse behavior.}

\subsection{$RQ_2$: How do producers document PTLMs, and how do these practices align with consumer information needs?}

\begin{figure*}[t]
\centering
\includegraphics[width=\linewidth]{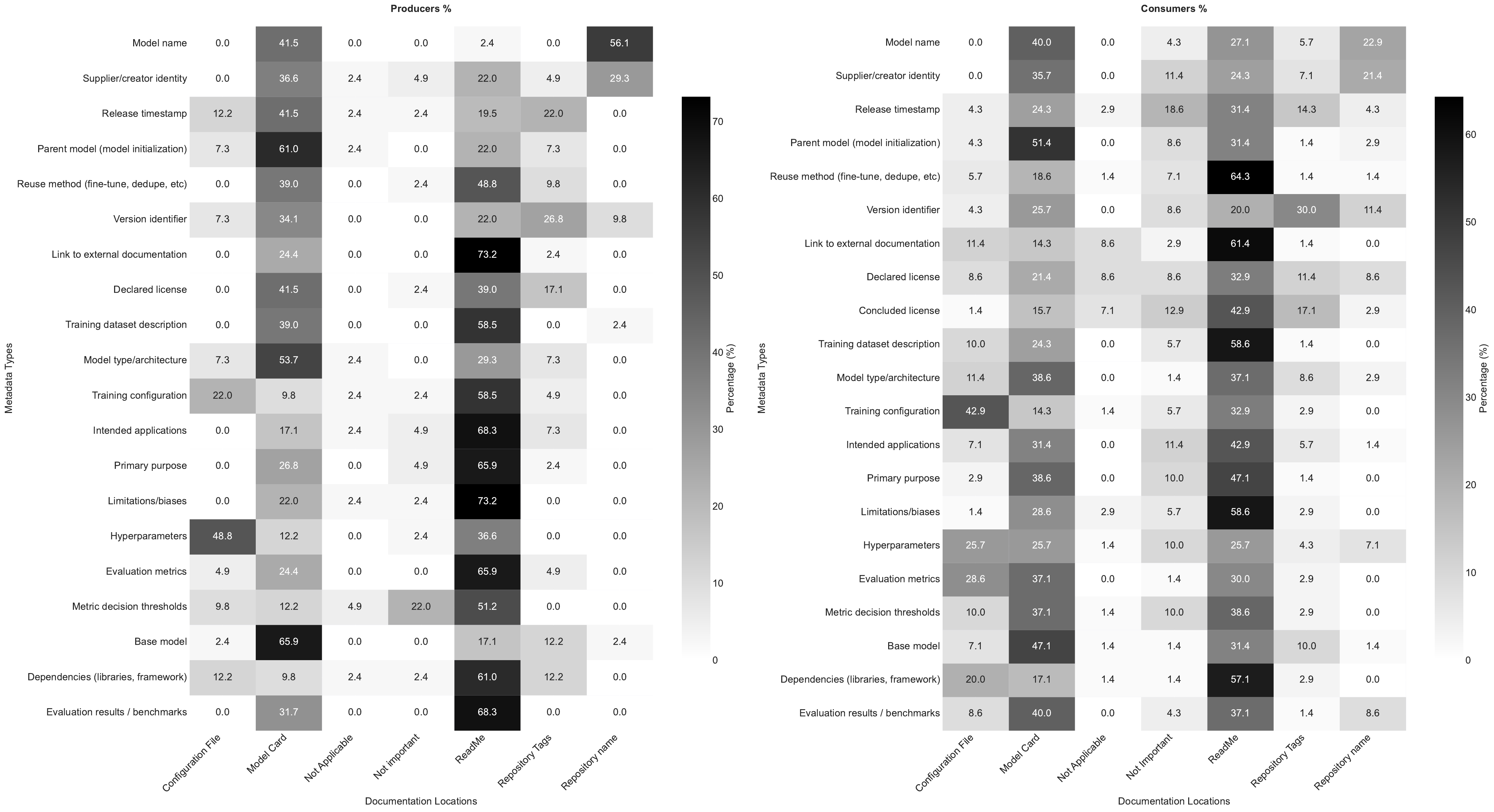}
\caption{\textbf{Primary documentation locations for various model metadata types, comparing producers and consumers. The values represent the percentage distribution of responses for each metadata category.}}
\label{doc_heatmap}
\end{figure*}

\noindent\textbf{\Producers and \consumers agree on primary model documentation artifacts but systematically diverge on where specific metadata types belong.} For instance, \producers predominantly encode model names within their HF repository's name (56.1\%), while \consumers expect them in Model Cards (40.6\%) (\Cref{doc_heatmap}. Furthermore, \producers document training configurations in README files (67.5\%), while \consumers expect them in training configuration files (42.0\%) to reproduce them more easily. Furthermore, producers also place evaluation metrics in README files (65.0\%), while consumers expect them in Model Cards (37.7\%), which are markdown files within the same repository. Although the information is present, its placement in unexpected locations creates systematic friction. Producers aim to release models quickly with minimal overhead, while consumers want to reproduce results and validate models efficiently. Notably, for 16 out of the 20 types of model information for which we asked each cohort where to store or find it, the survey responses were statistically significantly different (separate Chi-square test per information type with medium to large Cramér V effect sizes). This suggests that the documentation misalignment is systematic rather than due to chance. This adds an interesting nuance to older studies that documented poor documentation practices on Hugging Face~\citep{stalnaker2025empirical, ajibode2025towards, kadasi2025model}, as even when documentation exists, it seems to be scattered across different locations within the same repository, indicating that improved documentation alone is insufficient.

\noindent\textbf{Producer documentation is often incomplete due to time pressure and technical complexity.} Only 22.0\% of producers report providing complete documentation for 81--100\% of their released models, while 54.0\% do so for at least 41\% of their models. When releasing models, producers cite time pressure (65.7\%) and technical complexity (40.0\%) as primary documentation barriers, while competitive disadvantage (e.g., helping rivals) (13.2\%) and proprietary concerns (e.g., base model details being proprietary) (10.5\%) are the least common in \Cref{documentation_barriers}. This behavior aligns with broader open-source trends where upstream open-source developers face severe time constraints \citep{steinmacher2014preliminary} and project documentation is rarely updated completely \citep{seshadri2024omnibor}.

\begin{table}[t]
\centering
\caption{Factors Preventing Complete Model Documentation among Producers}
\label{documentation_barriers}
\resizebox{\linewidth}{!}{%
\begin{tabular}{lrrrrrrrr}
\hline
\textbf{Barrier} & \textbf{Alw. (\%)} & \textbf{Oft. (\%)} & \textbf{Som. (\%)} & \textbf{Rar. (\%)} & \textbf{Nev. (\%)} & \textbf{Mean} & \textbf{A/O (\%)} & \textbf{Wtd. (\%)} \\ \hline
Time pressure / deadlines & 20.5 & 43.6 & 17.9 & 12.8 & 5.1 & 3.62 & 64.1 & 14.4 \\
Technical complexity of capturing all details & 2.6 & 35.9 & 23.1 & 17.9 & 20.5 & 2.82 & 38.5 & 11.2 \\
No clear standard for what to include & 5.1 & 23.1 & 30.8 & 28.2 & 12.8 & 2.79 & 28.2 & 11.1 \\
Documentation feels unnecessary for this model & 2.6 & 12.8 & 41.0 & 23.1 & 20.5 & 2.54 & 15.4 & 10.1 \\
There is little demand for detailed documentation & 2.6 & 25.6 & 12.8 & 35.9 & 23.1 & 2.49 & 28.2 & 9.9 \\
Dataset details are proprietary & 2.6 & 10.3 & 28.2 & 35.9 & 23.1 & 2.33 & 12.8 & 9.3 \\
Uncertainty about safe or appropriate disclosure & 5.1 & 12.8 & 15.4 & 41.0 & 25.6 & 2.31 & 17.9 & 9.2 \\
Available tools do not adequately support doc. & 5.1 & 7.7 & 17.9 & 43.6 & 25.6 & 2.23 & 12.8 & 8.9 \\
Base model details are proprietary & 5.1 & 5.1 & 15.4 & 38.5 & 35.9 & 2.05 & 10.3 & 8.1 \\
Competitive concerns discourage sharing info. & 10.3 & 2.6 & 5.1 & 41.0 & 41.0 & 2.00 & 12.8 & 7.9 \\ \hline
\end{tabular}%
}
\smallskip
\small\\ Note: Frequency column percentages (Always to Never) are calculated row-wise per factor. \textit{Mean} scores range from 1 (Never) to 5 (Always). \textit{Always/Often (A/O \%)} combines the top two frequencies. \textit{Weighted Percentage (Wtd. \%)} reflects a factor's relative share of total friction points across all items.
\end{table}

\noindent\textbf{Consumers compensate for incomplete documentation through workarounds, including empirical testing, community engagement, and model abandonment.} These documentation practices coincide with our observation that consumers rely on external repositories, community discussions, and empirical testing during model selection. Among consumers, 51.4\% remain neutral regarding the ease of model selection, 40\% find it easy or extremely easy, and only 8.6\% explicitly report it as hard. The 40\% who find selection easy despite low documentation completeness are those who have developed effective workarounds, not those who actually find the documentation adequate. 

The open question result in \Cref{all_open_text} corroborates these observations, showing that when documentation is missing, consumers compensate via cross-platform triangulation \& search, empirical verification \& sandbox testing, community engagement, and model switching. If a given model is unsuccessful, they would abandon it and consider a different model (given the large number of models available on platforms like Hugging Face). A typical example of how one consumer compensates through cross-platform triangulation and search: \textit{``I check alternative sources such as the model's GitHub repository and community discussions to infer missing details. If clarity is still lacking, I perform small-scale experiments, run profiling to identify undocumented memory and latency constraints, or switch to a better-documented alternative."} While prior SE documentation studies identify missing or incomplete documentation as a structural problem \citep{robillard2011field, aghajani2019software}, our survey explicitly links producers' model publishing barriers to these consumer workarounds, capturing the underlying human operational decisions.

Unlike traditional software development where documentation gaps can slow onboarding, insufficient PTLM documentation creates a distinct dynamic in the AI supply chain: consumers spend considerable time in the model selection process, and if uncertainty persists, they often abandon the model entirely and switch to an alternative. As one consumer clarified: \textit{``If critical information is missing, I usually check the linked repository or papers. If the uncertainty remains, I avoid using the model for production-like reuse or treat it only as an experimental baseline.''} Unlike traditional libraries where API behavior is deterministic and well-specified, PTLM behavior is stochastic and depends on undocumented training data and hyperparameters, making empirical testing and abandonment more common when documentation is insufficient.

\takeaway{Summary of Findings}{Producers reported that time pressure and technical complexity contribute to incomplete and misaligned PTLM documentation, while consumers reported compensating through cross-platform searches, testing, and community signals.}

\vspace{-0.9em}
\subsection{$RQ_3$: What practical strategies do producers and consumers use to understand and trace model lineage across the AI supply chain, and what limitations or challenges do these strategies exhibit?}
\noindent\textbf{Model lineage tracing remains rare, and producers and consumers trace for different purposes when they can.} Only 27.9\% of producers and 31.4\% of consumers have been able to trace the lineage of models they are aiming to reuse. This low rate does not reflect lack of interest, but stems from difficulties due to the documentation gaps identified in RQ2: without complete and structured metadata, producers and consumers cannot reconstruct a model's provenance even when they want to. When tracing does occur, \Cref{all_open_text} shows that \producers trace primarily for legal compliance and lineage validation while \consumers trace primarily for legal compliance, data bias awareness, and reliability evaluation.

One producer illustrated this perspective: \textit{``Understanding PTLM lineage is very important; modern models are rarely built from scratch; they are fine-tuned, quantized, merged, or distilled from an original base model. Knowing where the model came from and all along its way helps in technical stability, legal compliance, and performance optimization.''} A consumer added: \textit{``It is necessary to analyze whether the PTLM is suitable for the project. Even if the performance is good, it will be excluded from application if it is not suitable.''}

In other words, Hugging Face producers look ``backward" to verify provenance and compliance of the model they use as the basis for their own model, while GitHub consumers look ``forward" to assess deployment suitability within their specific context. 

\noindent\textbf{When tracing lineage, both groups rely on documentation and metadata inspection, though consumers also use multi-source verification, while verification of the base model relies on metadata and reputation rather than technical inspection.} Both \producers and \consumers employ four groups of approaches to trace model lineage beyond the immediate parent: documentation and metadata inspection, platform and tool-based tracing, repository and version tracking, and multi-source verification. However, as shown in \Cref{all_open_text}, the emphasis differs. Producers rely primarily on documentation examination and model configuration auditing, while consumers complement documentation examination by external cross-referencing. 

According to the closed survey responses, both groups most commonly verify the base model they work with using platform-provided metadata signals (e.g., repository tags and model cards) (\producers 60.0\%, \consumers 40.0\%), followed by academic publications (\producers 56.0\%, \consumers 34.7\%) and organizational reputation (\producers 48.0\%, \consumers 24.2\%). Direct technical verification through manual audit or weight inspection is less common (\producers 24.0\%, \consumers 14.7\%).

This reliance on surface-level signals underscores that while producers and consumers depend on metadata that is often incomplete or missing, they lack the tools or incentives to perform deeper verification. For most, lineage analysis remains aspirational. Even though 13.6\% of producers and 14.3\% of consumers report using lineage tooling, they report that these tools are not robust and cannot automate parts of the process due to heterogeneous information spread across platforms, inconsistent naming, and lack of semantic versioning.

\noindent\textbf{Model registry fragmentation and undocumented base models are the most critical lineage tracing barriers for both groups.} Both producers and consumers identify scattered information across platforms (producers 54.0\%, consumers 42.1\%) and undocumented base models (producers 40.0\%, consumers 45.3\%) as the most critical barriers. There are several additional obstacles. For instance, the length of model lineage chains can be non-trivial, since often multiple subsequent passes of fine-tuning, merging, or compression have been applied onto a base model like Llama-3, where intermediate modifications (and their metadata) become obscured and inseparable (producers 36.0\%, consumers 24.2\%). Furthermore, traceability from downstream models back to upstream parents is often missing (producers 36.0\%, consumers 22.1\%) and there is also a lack of standard file format for capturing lineage information (producers 34.0\%, consumers 27.4\%). 

\begin{table}[t]
\centering
\caption{Model Lineage Tracing Obstacles: Producers vs. Consumers}
\label{tab:lineage_obstacles}
\resizebox{\linewidth}{!}{%
\begin{tabular}{lrr}
\hline
\textbf{Obstacle} & \textbf{Producer (\%)} & \textbf{Consumer (\%)} \\ \hline
Information exists but is scattered across platforms (GitHub, HF, papers) & 54.0 & 42.1 \\
Base model not documented by creators & 40.0 & 45.3 \\
Multiple overlapping adaptations (hard to disentangle) & 36.0 & 24.2 \\
Downstream models are not clearly linked to their upstream models & 36.0 & 22.1 \\
No standard format for lineage information & 34.0 & 27.4 \\
Inconsistent model naming makes lineage difficult to trace & 26.0 & 23.2 \\
Configuration files don't consistently include base model information & 22.0 & 20.0 \\
Other & 0.0 & 2.1 \\ \hline
\end{tabular}%
}
\smallskip
\small\\ Note: Respondents could select multiple options. Percentages represent the proportion of respondents within each cohort who identified the respective factor as an obstacle.
\end{table}

Prior work on software supply chain traceability identifies missing dependency information as a core risk \citep{stalnaker2025empirical, srinivasan2025bomfather}, even though in traditional software ecosystems like Python and Node.js, there are standardized supply chain ecosystem such as PyPI and npm. The PTLM supply chain lacks such uniformity, since model provenance is fragmented across Hugging Face, GitHub, and academic papers, with no enforcement of complete metadata. Our study shows that this model registry fragmentation is perceived as equally problematic as having no documentation at all.

\noindent\textbf{When a parent or base model is suspected to have a bias or security flaw, producers evaluate the flaw through dependency tracing \& impact analysis, while consumers prioritize mitigation \& remediation actions, in addition to empirical testing \& evaluation being commonly used by both cohorts.} When a producer discovers that a model they built upon contains a bias or security flaw or a consumer does so for a PTLM they are using in their application, \Cref{all_open_text} shows that both use four groups of methods to identify which of their own models or applications are affected. 

Although both groups share a strong focus on behavioral assessment, producers also favor lineage tracking and component auditing, while consumers favor risk classification, remediation verification testing, version-level vulnerability profiling, and to a lesser extent, model configuration analysis. Component auditing and model retraining assessment are less commonly used by producers, while component isolation diagnostics and differential output analysis are only used by consumers. 

Overall, both groups rely more on model behavioral assessment than on systematic dependency analysis when responding to flaws. This contrast is reflected in participant responses: a producer stated \textit{``run targeted evaluations on my model,''} while a consumer described a broader response workflow: \textit{``Not willing to reveal trade secrets, sorry. High level, we utilize containment, classification, evaluation, and ultimately change which model if needed.''} 

Unlike established software supply chains where automated dependency discovery and visualization tools are mature~\citep{srinivasan2025bomfather}, the AI supply chain currently lacks comparable infrastructure for tracing model lineage. Consistent with this, our participants reported relying on ad-hoc empirical testing rather than structured dependency analysis when flaws were discovered. Prior work on software supply chains shows that tracking all dependencies is critical to mitigating supply chain attacks~\citep{seshadri2024omnibor}, yet our findings suggest that PTLM practitioners rarely apply comparable structured approaches.

\takeaway{Summary of Findings}{PTLM lineage tracing remains rare due to incomplete documentation, not lack of interest. When tracing occurs, producers seek legal compliance while consumers seek reliability evaluation. Both groups rely on documentation inspection, though consumers also use external cross-referencing. Fragmented registries and undocumented base models are the most critical barriers.}

\vspace{-5pt}
\subsection{$RQ_4$: What governance mechanisms do \producers and \consumers prioritize, and what barriers hinder their adoption?}

\noindent\textbf{Producers prefer mechanisms that can improve operational efficiency and version standardization, while consumers prefer documentation and transparency mechanisms as potential improvements for the AI supply chain.} Despite the descriptive differences in the level of reported influence, both producers (80.0\%) and consumers (61.1\%) generally view governance mechanisms as important factors in model reuse decisions. \Cref{all_open_text} shows that producers selected Standardized Model Naming Convention (applying naming rules for models), Enforced Semantic Versioning (mandatory version numbering that conveys the nature of changes in a new model version), and Visual Lineage Graph (graphical visualization of parent-child derivation chains) as their top three mechanisms. Consumers selected Mandatory Model Card Template (required structured documentation for each model), Dependency Disclosure Requirement (explicit declaration of parent models and training dependencies), and Standardized Model Naming Convention. 

\Cref{all_open_text} shows that the reasons behind these choices depend on the role. Producers who selected Standardized Model Naming Convention believed it would reduce confusion about what each model represents and where it came from. One producer explained this challenge: \textit{``Having a clear naming logic means I can stop bugging my coworkers to explain where they stored the latest model weights.''} Consumers also saw this mechanism as important, believing it would help with model repository uniformity (i.e., consistent naming and organization across repositories). As one consumer put it: \textit{``The naming is the most important one, IMO, as everyone sees the name and remembers it to an extent, the other information is more likely to be lost for other people, that would e.g. see the usage of the PLTM.''} Producers who selected Enforced Semantic Versioning believed it would help with model lifecycle tracking and model repository uniformity, while those who selected Visual Lineage Graph believed it would help with lineage auditing and friction reduction.

Similar to the reasons given by the producers for their governance mechanisms, consumers who selected Mandatory Model Card Template again believed it would enhance model repository uniformity. One consumer stated: \textit{``Directly addresses the most common pain point: incomplete documentation. A required template ensures at minimum dataset provenance, intended use, and limitations are covered before publishing. Low technical complexity, high practical impact. HF already has a model card format, enforcing it is a small step.''} Consumers who selected Dependency Disclosure Requirement believed it would promote model trustworthiness and help with lineage auditing. Finally, consumers who selected Standardized Model Naming Convention also believed it would help with model repository uniformity. 

The misalignment between producers and consumers is evident in how each group prioritizes versioning: producers view strict versioning as a way to maintain consistency during model deployment and to prevent unexpected changes across model registries, while consumers, though still benefiting from versioning information to assess update risks, place greater emphasis on other mechanisms. As one consumer explained: \textit{``Everything other than Dependency Disclosure Requirement doesn't really matter and is difficult to enforce. The implementation and security teams should be able to audit the model regardless of high level details being available or missing.''}

\noindent\textbf{Producers and consumers both consider Mandatory Adoption of the SPDX 3.0 AI Profile (producer 60\% vs consumer 34.7\%), Single Hugging Repository Requirement (producer 42.0\% vs consumer 26.3\%), and AI Bill of Materials (AIBOM) Generation Tool (producer 32\% vs consumer 20\%) as the mechanisms most likely to face adoption resistance.} These mechanisms were previously not prioritized by the cohorts in Fig. 2, and the resistance percentages are from our replication package. 

Regarding the idea to consolidate all model versions and checkpoints into a single Hugging Face repository (Single Repository Requirement), both groups believe that this will cause workflow disruption. A producer explained that \textit{``requiring consolidation [of all model versions and checkpoints] into a single repo disrupts existing workflows where separate repos per [model] checkpoint or [model] version are standard practice and serve different purposes,''} while a consumer argued that \textit{``different versions often have genuinely different purposes, licenses, or maintainers; forcing consolidation goes against how open-source development naturally works on platforms like HF and GitHub.''}

For the utilization of automated tools that can generate AIBOMs, producers believe that it will cause implementation overhead due to the difficulty of tracking AIBOM information from every data source and library, while consumers are concerned about tool unreliability and the risk that inaccurate outputs would create more problems than they solve. In fact, in \Cref{all_open_text}, no surveyed producer identified AIBOM generation tools as a potential improvement for AI supply chain governance, while some consumers did demonstrate that it might be useful for producers. A producer described the burden of traceability at scale, noting that \textit{``trying to track every single tiny data source and library for a complex model is like trying to count every grain of sand on a beach,''} whereas a consumer observed that \textit{``most tools generate hallucinated or incomplete data because they can't actually see inside the model, and without very high accuracy, it would create more problems than it solves.''}

Additionally, none of the consumers selected Configuration Schema Validation as either beneficial or difficult to implement. Similarly, none of the producers selected Platform-Level Badges as difficult to adopt, while no consumers selected Mandatory Adoption of the SPDX 3.0 AI Profile as beneficial. Considering the specific resistance patterns observed across these mechanisms, the respondents do not disagree about the value of governance mechanisms, yet they are concerned about integration into their existing engineering workflows.

\takeaway{Summary of Findings}{\Producers prioritize mechanisms that streamline model release and management, while \consumers prioritize mechanisms that improve transparency and model selection. Both groups agree on which mechanisms face the greatest adoption resistance, with model reuse and compatibility challenges serving as the primary barrier to adoption rather than disagreement over their value.}

\vspace{-10pt}
\section{Discussion and Implications}
% \vspace{-10pt}
Our findings suggest that producer-consumer misalignment is a useful lens for understanding AI supply chain challenges identified in repository mining studies. Beyond missing metadata or incomplete lineage, we observe persistent differences in producer and consumer priorities. Producers prioritize efficient model release, while consumers prioritize efficient evaluation and reuse, reflecting a recurring trade-off where reduced producer effort aligns with increased consumer effort.

\subsection{Producer and Consumer Priorities Create Different Integration Experiences}

A central pattern in our findings is that producers and consumers experience the same supply chain differently. Producers reported that time pressure and technical complexity influence where and how they document model information, while consumers preferred more structured locations for the same information. This asymmetry is evident across several RQ2 findings. Rather than requesting missing metadata, consumers adapt their evaluation process by searching GitHub repositories, inspecting associated papers, consulting community discussions, and running small-scale experiments. One consumer described this workflow: \emph{"I check the Hugging Face repo and GitHub (issues, discussions) for missing info. If still unclear, I run small local tests to verify behavior myself. As a last resort, I ask AI assistants or community forums (Discord, Reddit) for guidance."}

The same asymmetry appears in reported risks associated with model selection (RQ1). Producers highlight model issues such as data contamination and hidden bias, while consumers emphasize documentation quality, reproducibility, and performance concerns affecting integration. Release practices that simplify publication for model producers are thus associated with additional validation work for consumers, including repository inspection, empirical testing, and community consultation.

\noindent\textbf{Implication:} In the short term, platforms should make Model Cards mandatory through template-based documentation generators or validation checks that flag misplaced critical information. In the longer term, researchers should explore ways to automatically generate parts of model documentation from training scripts and configuration files, reducing the burden on producers while improving documentation completeness for consumers.

\subsection{Fragmented Provenance Limits Both Lineage and Governance}
The RQ1 and RQ2 findings on incomplete and fragmented model documentation indirectly relate to the adoption issues for model lineage tracing and supply chain governance encountered in RQ3 and RQ4. Approximately one third of respondents reported tracing model lineage beyond the immediate parent model, yet both producers and consumers identified scattered information across platforms and undocumented base models as the primary barriers. Producers trace lineage for provenance and compliance, consumers for reliability and application suitability, but both depend on the same fragmented sources. As one producer explained: \emph{"The supply chain for AI is so opaque that we often don't even know what is in the base models we are building on top of."}

Recent efforts such as SPDX 3.0 AI Profile \citep{bennet2025implementing}, automated AIBOM generation \citep{vandendriessche2026aibomgen}, and tools such as model provenance kit\footnote{https://github.com/cisco-ai-defense/model-provenance-kit} aim to improve provenance visibility, but depend on the same metadata that practitioners already report as incomplete or inconsistently organized. One consumer noted the practical consequences: \emph{"The generation process is buggy. It crashes as soon as it hits non-standard data types."} As a consequence, practitioners rely on lower-effort alternatives such as organizational reputation, platform metadata, or empirical testing. Even the 11.6\% of consumers who perform no lineage verification illustrate this trade-off, consciously substituting institutional trust for technical verification to maintain development speed.

This, in turn, led to both groups consistently associating adoption resistance with additional workflow effort. This pattern appears for mechanisms such as Mandatory Adoption of the SPDX 3.0 AI Profile, Single Repository Requirement, and AI Bill of Materials Generation Tool. Producers emphasized difficulty collecting and organizing provenance information, while consumers emphasized implementation overhead and limited practical benefit given current metadata quality. Although the AIBOM generation tool mechanism can be largely automated after model training \citep{vandendriessche2026aibomgen}, the provenance information for parent or base models must first exist in a complete and structured form. Our findings suggest that this effort, rather than the generation process itself, is the primary obstacle.

\noindent\textbf{Implication:} Governance mechanisms should integrate with existing development workflows rather than introduce separate documentation activities. For producers, this means capturing provenance information during model development in machine-readable formats, similar to how SBOMs are generated during CI builds. For consumers, platforms could reduce effort through lineage visualization and cross-platform provenance aggregation, instead of requiring manual searches across repositories, documentation, and publications. If researchers find ways to more reliably create AIBOMs by linking and analyzing distributed sources of lineage information, this could help AIBOM adoption could improve, making that practice more viable for practitioners.

\section{Threats to Validity}\label{threats}
\noindent\textbf{External Threats:} A key threat is the modest response rates (7.17\% for Hugging Face and 5.07\% for GitHub) and sample size of 145 respondents. Low response rates are a well-known challenge in software engineering surveys~\citep{baltes2022sampling}, particularly when contacting open source contributors through cold email with no prior relationship or incentive. Despite this, our findings remain credible as respondents spanned diverse roles and experience levels, qualitative responses reached saturation, and results align with repository mining studies that identified related documentation and versioning challenges through different methods~\citep{stalnaker2025empirical, ajibode2025towards}. Email tracking indicated that 45\% (producers) and 47\% (GitHub) of invitations were not opened, meaning that the effective response rate among those who saw the invitation was higher than the nominal rate suggests.

Additionally, our recruitment strategy combined probability-based sampling (Phase 1) with purposive sampling of recently active contributors (Phase 2). As a result, our findings reflect the perspectives of reachable and active practitioners and may not capture the full diversity of the global PTLM supply chain. Furthermore, because our data collection focused on the two most widely used open-source platforms for PTLM development and model reuse, Hugging Face and GitHub, these findings may not fully generalize to proprietary PTLM supply chains. Corporate or closed-source environments often operate under different organizational mandates, standardized tooling, and compliance pressures that could alter how producers and consumers interact.

\noindent\textbf{Internal Threats:} Modest response rates might introduce non-response bias, where practitioners facing severe model reuse failures might disproportionately participate. However, 37.9\% of \consumers maintained a strictly neutral stance regarding model selection ease (from RQ2), indicating that our sample includes practitioners who have normalized these supply chain deficits into routine workflows rather than consisting solely of those with extreme experiences. To mitigate remaining bias, we focus on reporting observed differences between groups rather than generalizing to the entire population, leaving telemetry-based validation for future work. As with all survey-based studies, responses reflect self-reported practices, which may not precisely correspond to actual behavior, a limitation inherent to the survey method~\citep{baltes2022sampling}.

\section{Conclusion}
We investigated how model producer practices compare to consumer needs across the AI supply chain through a survey of 50 Hugging Face producers and 95 GitHub consumers. Our findings reveal misalignments between model publication practices and model reuse needs, demonstrating that challenges extend beyond individual artifacts to reflect broader gaps in documentation standards, lineage visibility, and governance adoption. Our analysis shows that metadata is frequently misplaced rather than missing. Furthermore, lineage tracing beyond the immediate parent model remains rare, practiced by under 32\% of respondents. Finally, producers favor governance mechanisms that streamline release processes, while consumers prioritize model cards and dependency disclosure. Although AIBOM mechanisms like that of SPDX 3.0 offers AI profiles that could address many of these needs, practitioners resist adoption due to limited awareness, implementation challenges, and workflow friction. Overall, our study takes an important first step toward understanding the human processes underlying these supply chain issues and suggests that future work should focus on aligning producer incentives with consumer information needs and improving the usability of governance mechanisms through better workflow integration.

\section*{Acknowledgement}
We thank the 145 producers and consumers who participated in our survey for their time and valuable insights.

\section*{Data Availability}
\label{sec:availability}
The datasets generated and analyzed during this study are available in the replication package~%\citepp{SemFin}.
\section*{Funding} 
This research was supported by the NSERC Discovery Grant RGPIN-2025-04654.
\section*{Ethical Approval} This study does not involve human participants or animals.
\section*{Informed Consent} No human subjects were involved in this study.
\section*{Conflicts of Interests/Competing Interests}
The authors declare that they have no known competing interests or personal relationships that could have (appeared to) influenced the work reported in this article.
\section*{Author Contributions}
\begin{itemize}
    \item Adekunle Ajibode: Conceptualization, Data Collection, Methodology, Data Analysis, Writing – Original Draft.
    \item Oussama Ben Sghaier: Methodology, Data Validation, Writing – Review \& Editing.
    \item Keheliya Gallaba: Research Direction - Review \& Editing.
    \item Bram Adams: Supervision, Writing – Review \& Editing, Conceptual Guidance, Research Direction.
    \item Ahmed E. Hassan: Supervision, Research Direction.
\end{itemize}

\bibliographystyle{plainnat}
\bibliography{main_draft}

\end{document}